\documentclass[letterpaper,twocolumn,10pt]{article}
\usepackage{usenix2019_v3}

\usepackage{xurl}
\usepackage{tabto}
\usepackage{tikz}
\usepackage{amsmath}
\usepackage{nicefrac}
\usepackage{siunitx}
\usepackage{array,framed}
\usepackage{booktabs}
\usepackage[htt]{hyphenat}
\usepackage{
  color,
  float,
  epsfig,
  wrapfig,
  graphics,
  graphicx,
  subcaption
}
\usepackage{textcomp,amssymb}
\usepackage{setspace}
\usepackage{latexsym,fancyhdr,url}
\usepackage{enumerate}
\usepackage{algorithm2e}
\usepackage{algpseudocode}
\usepackage{graphics}
\usepackage{xparse} 
\usepackage{xspace}
\usepackage{multirow}
\usepackage{csvsimple}
\usepackage{balance}

\usepackage[labelfont=bf]{caption}
\usepackage{tabularx}
\usepackage{float}

\usepackage{array}
\newcommand{\PreserveBackslash}[1]{\let\temp=\\#1\let\\=\temp}
\newcolumntype{C}[1]{>{\PreserveBackslash\centering}p{#1}}
\newcolumntype{R}[1]{>{\PreserveBackslash\raggedleft}p{#1}}
\newcolumntype{L}[1]{>{\PreserveBackslash\raggedright}p{#1}}

\begin{document}

\raggedbottom

\date{}

\title{\Large \bf KeyDroid: A Large-Scale Analysis of\\ Secure Key Storage in Android Apps}

\author{
{\rm Jenny Blessing}\\
University of Cambridge
\and
{\rm Ross J. Anderson}\\
University of Cambridge\\
University of Edinburgh
\and
{\rm Alastair R. Beresford}\\
University of Cambridge
} 

\maketitle

\begin{abstract}
Most contemporary mobile devices offer hardware-backed storage for cryptographic keys, user data, and other sensitive credentials. Such hardware protects credentials from extraction by an adversary who has compromised the main operating system, such as a malicious third-party app. Since 2011, Android app developers can access trusted hardware via the Android Keystore API~\cite{keystore_official}.
In this work, we conduct the first comprehensive survey of hardware-backed key storage in Android devices. 
We analyze 490,119 Android apps, collecting data on how trusted hardware is used by app developers (if used at all) and cross-referencing our findings with sensitive user data collected by each app, as self-reported by developers via the Play Store’s data safety labels~\cite{google_dev_data_safety}.

We find that despite industry-wide initiatives to encourage adoption, 56.3\% of apps self-reporting as processing sensitive user data do not use Android’s trusted hardware capabilities at all, while just 5.03\% of apps collecting some form of sensitive data use the strongest form of trusted hardware, a secure element distinct from the main processor.
To better understand the potential downsides of using secure hardware, we conduct the first empirical analysis of trusted hardware performance in mobile devices, measuring the runtime of common cryptographic operations across both software- and hardware-backed keystores. We find that while hardware-backed key storage using a coprocessor is viable for most common cryptographic operations, secure elements capable of preventing more advanced attacks make performance infeasible for symmetric encryption with non-negligible payloads and any kind of asymmetric encryption.
\end{abstract}

\section{Introduction}
\label{sec:intro}

Mobile devices store highly sensitive user data ranging from private health information and payment credentials to personal photographs and correspondence. At the same time, mobile handsets are regularly lost or stolen, making data stored on devices vulnerable to an adversary with physical device access. Similarly, sensitive data may be accessed by an adversary who is able to compromise the main operating system (OS), such as malicious third-party apps which circumvented app store vetting processes or were independently downloaded by users~\cite{zhou2012detecting,wang2014exploring}.

Modern encryption methods may provide data confidentiality and integrity against such threats, but data is only as secure as the cryptographic keys used. Keys stored in a software keystore (e.g., Java’s Bouncy Castle keystore) are vulnerable to memory-extraction attacks~\cite{tang2012cleanos,maartmann2009persistence} where an adversary with full control over the operating system or physical access to the device can retrieve the decryption keys or other sensitive data through a memory dump of device RAM.

Fortunately, mobile device key storage has seen major security improvements over the past decade: almost all modern mobile handsets now offer some form of hardware-backed credential storage capable of protecting keys against an adversary with root permissions~\cite{hugenroth2023sloth}. The most common form of hardware-backed storage (commonly called ``trusted hardware'' or “secure hardware”) is the trusted execution environment (TEE), a special mode of operation by the main processor (e.g., Arm TrustZone or Intel VT). Keys are generated and stored within specialized hardware, and all cryptographic operations using these keys take place within the hardware component. Provided the TEE is not compromised, these operations cannot be inspected or interfered with by the Android OS (e.g., an attacker who compromises the device cannot extract keys or use them to decrypt data stored off-device). Android has offered the Android Keystore system~\cite{keystore_official} as its public trusted hardware API for developers since 2011.

Recent premium models of Android smartphones such as the Google Pixel devices contain additional hardware in the form of a separate secure processor, commonly known as a secure element (SE)~\cite{keystore_official,android_pixel6}. While a TEE is a separate OS on the main processor, an SE is an entirely separate processor with its own CPU, memory, and storage. In Android, the SE is called the \texttt{StrongBox Keymaster}~\cite{android_strongbox}. The Android Keystore API uses the device's TEE by default but offers developers the option of requesting StrongBox instead.

Unfortunately, there is currently a lack of empirical evidence on when and how developers use secure hardware in practice. Secure hardware is only useful if it is actually used, and the Android Security team acknowledges that apps need to explicitly use these APIs in order to see a security benefit as Android’s historical Java cryptography APIs use a software-backed keystore by default~\cite{mayrhofer2021android}. Furthermore, while hardware-backed keystores provide significant security benefits, runtime performance is a critical consideration for mobile developers. More advanced forms of secure hardware (e.g., StrongBox), tend to come with an accompanying performance hit, as acknowledged at a high level in Android's documentation~\cite{android_strongbox}, but to date there has been no publicly available empirical data on hardware keystore performance to the best of our knowledge. At the same time, however, Android is pursuing public initiatives to encourage wider adoption of secure hardware such as the Android Ready SE Alliance (see~\S\ref{sec:secure_element}). The lack of empirical evaluation of performance and existing usage patterns is a major barrier to encouraging more widespread adoption: without detailed performance statistics, developers cannot make informed choices about the trade-offs between security and performance for their use case.

In this work, we conduct the first comprehensive and systematic study of secure credential storage in Android, analyzing both the contemporary usage and performance of key storage schemes. We compile and analyze a dataset of 490,119 Android applications between October 2023 and August 2024, extracting data from 64 API calls relevant to key storage in Android. We find that 56.3\% of apps report collecting sensitive data as part of the Play Store's data safety labels do not use any form of trusted hardware, and only 5.03\% contain a reference to the SE API. Moreover, these usage figures represent an upper bound on security within the Android app ecosystem as it is not possible to detect at scale whether apps which contain at least one reference to the Android Keystore API are using it to secure all sensitive and relevant credentials. In particular, of those apps that do use the Android Keystore API, 94.7\% of key initializations are located in third-party components, indicating that use of the Keystore API may be due to using a general-purpose library rather than a conscious choice to use hardware-backed key storage. Furthermore, we find that 8.5\% of keys generated in the Android Keystore explicitly disable Android’s randomized encryption requirement (i.e., IND-CPA), indicating that secure defaults are not enough to enforce security guarantees.

Having measured the usage of secure key storage across the Android app ecosystem, we investigate the runtime performance of common cryptographic operations using the hardware-backed key storage APIs to consider whether performance overhead may discourage adoption. We find that the performance of TEE-backed key storage is viable for the vast majority of common app use cases and is noticeably different from a software-backed keystore only for large payloads greater than 5 MiB. StrongBox introduces a far more significant performance hit. For instance, encrypting a 1MiB message with AES-GCM takes around 3 seconds and simply \textit{generating} asymmetric keys in StrongBox takes over 9 seconds in Google’s flagship Pixel 8 device, a runtime which may be prohibitive even for security-conscious apps. Even so, StrongBox's performance has improved significantly since first introduced by Android in 2018 and is viable for use cases involving small payloads, such as using StrongBox to encrypt a key generated by a keystore with less overhead. To the best of our knowledge this is the first time comprehensive performance measurements of trusted hardware in mobile devices have been published, providing Android developers with empirical evidence to make informed decisions for their particular use case.

Our specific contributions are as follows:
\begin{itemize}
    \itemsep0em
    \item We design KeyDroid, a tool for static analysis of key storage in Android apps.
    \item Conduct large-scale static analysis of $\sim$500,000 apps to understand how trusted hardware is used, cross-referencing results with user data collection practices.
    \item Run comprehensive measurements of key storage performance on all hardware primitives in Android devices.
    \item Conduct a developer survey to better understand factors influencing developer decisions about trusted hardware.
    \item Provide developers with concrete guidance on trusted hardware usage patterns and performance.
\end{itemize}



\section{Key Storage in Android}
At a systems level, Android provides three options for storing cryptographic keys and other sensitive credentials: a software keystore via long-standing Java APIs, hardware-backed key storage logically separated from the main Android Operating System (OS) to protect against OS compromise, and hardware-backed key storage located on a separate processor to guard against the most advanced logical and physical attacks. 
We discuss security properties and limitations of each below.

\subsection{Software-backed Key Storage}
Mobile devices have historically relied on software keystores which operate within the mobile OS and use the device's internal storage.
Java's Cipher API~\cite{java_cipher} and the Java Keystore API~\cite{java_keystore} using a software-backed provider (either Bouncy Castle or AndroidOpenSSL, also known as Conscrypt~\cite{android_conscrypt}) are both examples of software-backed keystores within Android that have been available since Android's inception in 2007. On all Android devices today, if no keystore provider is specified when using Java's cryptographic APIs (namely \texttt{java.security.*} and \texttt{javax.crypto.*}) Android defaults to using a software-backed keystore even if the device supports hardware-backed key storage~\cite{android_cryptochanges,android_cryptography}.

Software key storage implementations are vulnerable to memory extraction attacks, where an adversary with root permissions in the Android operating system can observe the key as it is decrypted in RAM while being used~\cite{maartmann2009persistence,halderman2009lest}.
Mobile applications are particularly vulnerable to such attacks since apps are long-running processes and keys stored in memory are not garbage collected until a process has terminated. 
Malware, malicious third-party apps, and other privileged users are all capable of compromising the underlying OS, including kernel access control measures, and launching an attack of this sort.

Prior work investigating Java cryptography APIs has also observed that these libraries have an unfortunate tendency to use the weakest ciphers as defaults (ECB mode with symmetric encryption being the most pervasive example)~\cite{focardi2018mind,egele2013empirical}. Such choices shift the responsibility for achieving an adequate security level from the API provider to the developer.

\subsection{Hardware-backed Key Storage}
 The defining feature of hardware-backed key storage is that keys are stored and used in hardware separate from the main OS. A compromise of the Android operating system, then, will not compromise any cryptographic keys or other processes running inside the hardware element. Importantly, secure hardware ensures that keys will never be revealed in memory while they are used (and therefore cannot be viewed or extracted even by a privileged user).

The vast majority of modern smartphones today contain at least some form of secure execution environment~\cite{mayrhofer2021android,hugenroth2023sloth}. We use the terms ``secure hardware'', ``trusted hardware'', and ``hardware enclave'' interchangeably throughout this work to broadly characterize hardware-backed key storage.

There are two main forms of hardware-backed key storage in the Android ecosystem, each offering different security properties: (1) a trusted execution environment (TEE), available in Android through the \texttt{Android Keystore} API~\cite{keystore_official} and (2) a secure element (SE), termed (\texttt{StrongBox Keymaster}~\cite{android_strongbox} and provided as a subset of the Android Keystore API. We discuss each of these in turn below as different forms of secure hardware vary in degree of isolation from the Android OS, and hence the attacks they protect against. Throughout the rest of the paper, we use the terms ``Android Keystore'' and ``Keystore'' to refer specifically to Android's trusted hardware API (either TEE or SE).

\subsubsection{Trusted Execution Environment}
A trusted execution environment (TEE) is a discrete area of the main processor intended to provide a more secure, logically isolated execution environment. It has its own operating system (named Trusty in Android~\cite{android_trusty}), and communicates with the Android OS through requests forwarded through the Android Keystore interface to the TEE, referencing keys by a string alias. In Android, the TEE is located on the main processor, which is divided into the Android OS and the Trusty OS~\cite{android_trusty}, also commonly referred to as the \textit{normal world} and the \textit{secure world}. The hardware used to protect the normal world from the secure world depends on the processor architecture: TrustZone is used for ARM-based systems and provides dual execution environments~\cite{pinto2019demystifying} while Intel x86 uses virtualization technology to provide similar support~\cite{intel_virtualization}.

The primary security benefit of a TEE is to guard against kernel compromise, including malicious applications installed on the device which could request root permissions~\cite{mayrhofer2021android,sabt2016breaking}. The Android kernel and applications run in the normal world, while the secure world (i.e., the hardware enclave) stores long-term cryptographic key material and performs operations using these keys.
Trusted hardware has numerous other benefits for mobile device security, such as enabling hardware root of trust schemes to authenticate firmware running on the device, but in this paper we focus on the direct security to app developers for storing and using cryptographic keys.

In Android, a TEE has been available since Android 4.3 (API level 18) was released in 2013, with new features added over the years since~\cite{keystore_official,android_jellybean}.  
The initial version of the Android Keystore only supported asymmetric cryptographic operations and did not add support for symmetric keys until Android 6.0 (API level 23) in 2015 (approximately two years after the initial release date).
Hardware-level key attestation, the ability to verify that keys are indeed stored in a hardware-backed keystore, and other more advanced features were introduced in Android 7.0 (API level 24)~\cite{android_key_attestation}.

In addition to hardware-derived security benefits, the Keystore API makes deliberate design choices that provide an increased level of security in practice when compared with older Java APIs. The API explicitly disallows certain insecure key configurations, such as symmetric encryption with a constant initialization vector, and offers more secure defaults.

While TEEs offer substantial benefits over software-backed key storage, including protection from memory extraction, they are still vulnerable to various physical attacks, including side-channel attacks. 
There have been several documented attacks on Intel SGX~\cite{van2018foreshadow,chen2019sgxpectre,van2021cacheout}, which is an example of a TEE; most of these were side-channel attacks~\cite{nilsson2020survey}.
Prior work has also discovered several architectural design flaws in ARM TrustZone implementations leaving data stored even in TEEs potentially vulnerable to sophisticated threat actors~\cite{shakevsky2022trust,cerdeira2020sok}. 
To protect against the most advanced attacks, a device needs to contain a hardware element entirely separate from the main processor: a secure element.

\subsubsection{Secure Element}
\label{sec:secure_element}

Most premium Android smartphones include a secure element (SE), a form of hardware security module (HSM) which must have its own CPU and storage, tamper-resistant packaging, and a true random number generator~\cite{keystore_official,android_strongbox}. An SE provides all the benefits of a TEE and more: the increased isolation from the main Android OS and processor provides resistance to various side-channel attacks, including cold-boot memory attacks and shared-resource attacks~\cite{keystore_official}. As with a TEE, cryptographic keys are generated and stored within the confines of the SE, and any operations performed using the key material take place within the hardware so the key never enters an application's host memory. Different hardware elements are not mutually exclusive---for instance, a mobile device containing an entirely separate SE will almost certainly also contain a TEE as part of its main processor.

Android's public SE API is termed the \texttt{StrongBox Keymaster}~\cite{android_strongbox} (henceforth abbreviated as StrongBox) and has been available to external developers since Android 9.0 (API level 28) was released in August 2018~\cite{keystore_official}. 
An SE was first introduced in Pixel devices, Google’s flagship device line, with the Titan M chip (Google’s in-house secure element processor) in the Pixel 3 in 2018~\cite{android_pixel3}.
This was upgraded to the Titan M2 chip beginning with the Pixel 6 in 2021~\cite{android_pixel6}. 
System-on-chip (SoC) hardware, or integrated secure elements (iSE), qualify as providing StrongBox support as long as they meet the requirements above~\cite{keystore_official}. 
In 2021, Google’s Pixel 6 introduced Google Tensor, a system-on-chip (SoC) that is isolated from the main processor but also has its own CPU and ROM, among other features~\cite{android_pixel6}; Google Tensor interfaces with the Titan M2 chip. 
While secure elements are still comparatively novel in mobile handsets, Hugenroth et al.~\cite{hugenroth2023sloth} estimated secure element availability in contemporary mobile devices and found that as of 2023, 96\% of iPhones and 45\% of Android devices offer some form of SE.
We expect these percentages will increase in future years as older devices are cycled out.

Spurred on by the advanced security properties SEs can provide, industry firms have invested significant resources into encouraging the development and adoption of HSM schemes: Google launched the Android Ready SE Alliance in 2021, a ``collaboration between Google and Secure Element (SE) vendors'' that aims to make discrete hardware-backed storage (e.g., StrongBox) ``the lowest common denominator for the Android ecosystem'' and to facilitate interoperability and consistency across secure element vendors within the Android ecosystem~\cite{android_ready_se,android_ready_se_blog}. We discuss recommended best practices and legal mandates in further detail in \S\ref{appendix:best_practices}.

Despite the industry shift towards SEs as the desirable and intended outcome, to the best of our knowledge there have been no prior studies on the usage or performance of this form of trusted hardware. This is of particular concern since SEs are widely acknowledged to reduce performance. Google's documentation in particular described StrongBox’s performance as ``a little slower and resource-constrained (meaning that it supports fewer concurrent operations) compared to TEE'', and recommends StrongBox for developers who ``want to prioritize higher security guarantees over app resource efficiency''~\cite{keystore_official}. Due to these performance drawbacks, the Android Keystore API is structured so that developers must explicitly opt in to using StrongBox even when the application is running on a device that contains a SE.

\subsection{Key Considerations}
Trusted hardware is not a panacea: although hardware-backed key storage prevents keys from being exported off-device or revealed in memory, the keys can still be \textit{used} on-device by an attacker with root privileges, a ``fundamental limitation'' of hardware-backed storage~\cite{cooijmans2014analysis,mayrhofer2021android}.
Even so, the adversary will only be able to decrypt data stored on the device which, depending on the application, may limit the damage they can cause if they are unable to use the keys to decrypt data stored \textit{off} the device (e.g., data stored on a remote server).
Additional authentication requirements prior to key use can also substantially mitigate this risk.

Furthermore, the use of hardware-backed protection for cryptographic key material is ``best effort'' in the sense that the Android Keystore API uses the TEE if it is available on the device (or SE if specially requested), but reverts to a software-backed keystore otherwise.
The default reversion to a software-backed keystore instead of throwing an error reflects a desire to support backwards compatibility and a fragmented Android ecosystem containing many different device vendors with different price budgets and hardware specifications. Developers who desire to require hardware-backed storage as the minimum security level of their product can add runtime conditional checks hardware availability and adjust accordingly. A key goal of this work, then, is to explore whether app developers do indeed request to use trusted hardware on devices where it is available.

\begin{figure}
    \centering
    \includegraphics[width=1\columnwidth]{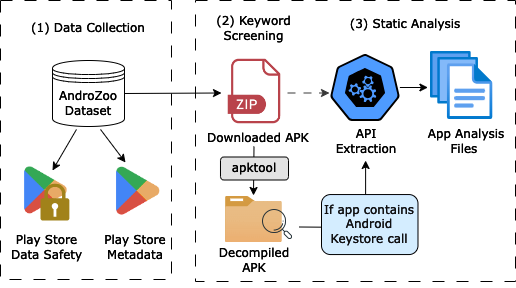}
    \caption{\textbf{KeyDroid Stages:} We (1) scrape Play Store metadata and data safety information for all apps in the AndroZoo dataset with at least 10,000 downloads and (2) decompile each app and pre-screen for any relevant API references. If an app contains a reference to the Android Keystore API, we run KeyDroid, our in-depth static analysis tool, to generate the app call graph and extract all API references, arguments, and call packages.}
\label{fig:keydroid}
\end{figure}

\section{Methodology}
\label{sec:methodology}

We begin by describing our process for collecting our dataset of Android apps and analyzing these apps with respect to API usage. Figure~\ref{fig:keydroid} provides a high-level overview of all app analysis stages. We further describe our methodology for testing the runtime performance of different keystores across common cryptographic operations.

\subsection{Dataset Selection}
We use the publicly available AndroZoo dataset~\cite{allix2016androzoo} as our source for Android applications. 
We initially identify 8,804,118 apps in the AndroZoo dataset from the Play Store marketplace which were crawled on or after July 2013, when Android’s trusted hardware API was first released to developers, though this number includes different versions of the same app and apps no longer available for download. 
We necessarily only consider free apps since the AndroZoo dataset does not include paid apps.

For each app provided in the AndroZoo dataset that passed preliminary filtering, we scrape the Play Store between October 2023 to March 2024 to filter for apps currently available at the point of scraping with at least 10,000 downloads.
We collect other relevant app metadata at the same time, resulting in a dataset of 490,119. 
We were able to successfully download and decompile almost all of these apps, leaving us with a revised dataset of 486,234.
We record the following metadata for each app: app package ID, title, number of installs, developer name and email, Play Store genre, release date of the latest version (release date of initial version is not available), and version number. 

We download the Android Package (.apk) archive file containing the app source code, metadata, and other resource files for each of these 486,234 apps. 
When there are multiple versions of the same app (as identified using Android’s APK package name) available in the AndroZoo dataset, we use the most recently crawled version.

\subsection{Play Store Data Safety Labels}
App key storage is only a concern if the app processes sensitive or confidential data. 
Since July 2022 Google has required each app listed in the Play Store to complete a data safety form containing self-reported information from the app developers on what types of user data the app collects and shares with third-parties, and for what purpose; this includes data collected by third-party libraries. 
For instance, the Signal messaging app notes that it collects only a user’s phone number for ``app functionality and account management'', and does not share data with third parties~\cite{signal_data_safety}. We provide further specifics on what data is considered to be sensitive and discuss limitations of developer-reported data in ~\S\ref{appendix:data_safety}.

\subsection{Static Analysis}
To reduce computational load, we use multiple layered static analysis techniques to filter for references to Android’s trusted hardware APIs and extract relevant API calls. We begin by executing a basic keyword search across all APKs in our dataset, and then perform more in-depth static analysis on any APKs flagged as relevant.

\label{sec:keyword_filtering}
\vspace{1em}
\noindent\textbf{Keyword Filtering.} Since analyzing the call graph is very resource-intensive, to filter candidate apps we first decompile each .apk file using apktool~\cite{apktool} and run an initial grep search for any call to the Android KeyStore API (\texttt{android.security.keystore}).
After filtering out any apps that do not contain at least one reference to the Keystore API, we are left with a dataset of 122,305 apps.

\label{interprocedural_call_graph_analysis}
\vspace{1em}
\noindent\textbf{Inter-Procedural Call Graph Analysis.}
To analyze the bytecode of the 303,948 apps flagged as having at least one relevant API call, we use Soot~\cite{soot}, a well-known framework for inter-procedural static analysis~\cite{li2017static} also used by similar related work. 
We experimented with using FlowDroid and other static analysis tools that more accurately model the Android lifecycle (e.g., by detecting implicit callback methods such as \texttt{onCreate} or \texttt{onClickListener}) but found that the runtime was sufficiently large as to make it infeasible for a dataset of our size, in large part due to its iterative callback calculation, which recomputes the call graph each time a new callback is encountered. 
Prior work~\cite{wu2021program} showed that Flowdroid did not finish app call graph generation on 24\% of apps even with a timeout of 5 hours, consistent with our own observations, and so we ultimately determined Soot offered the right balance of accuracy and efficiency.

We allocate each APK 10GB RAM and set an automatic timeout of 30 minutes. Our analysis tool begins by generating the call graph of the APK to determine the context for a particular API reference. To keep runtime manageable, we assume that all methods are reachable while generating the initial call graph, and conduct a custom reachability analysis (described in more detail below) tracing backwards from a method of interest along the method call chain.

We search for 64 distinct API calls, including all methods from the primary KeyStore API as well as other Android cryptography APIs that in turn call the KeyStore API, such as \texttt{androidx.security.crypto.MasterKey}~\cite{android_masterkey} and \texttt{android.security.keystore.KeyProtection}~\cite{android_keyprotection}, and the primary methods from the Java KeyStore~\cite{java_keystore} and Cipher APIs~\cite{java_cipher}.
The Java cryptographic APIs allow developers to specify a keystore provider, and so we check these to see if developers are referencing the Keystore API indirectly.
The full list of specific API methods searched for is available in our dataset in Table~\ref{sec:dataset} in the Appendix. For each API call identified, we collect the full method signature of the calling method, including associated package and class names, record the object on which the method is called (i.e., register value), and extract all parameter values by applying backwards program slicing~\cite{weiser1984program}. As part of our reachability analysis, we conduct a backwards breadth-first search and trace each method containing a relevant API call backwards through the call graph for up to 1,000 nodes, recording all possible paths.

\vspace{1em}
\noindent\textbf{Package Analysis.} We are particularly interested in determining whether a particular API call is located within the main application code or whether it is part of a third-party library. First-party usage indicates that developers have consciously chosen to store cryptographic key material in trusted hardware, while for certain third-party libraries developers may be unaware that this is even occurring.

To determine call context, we classify packages as first- or third-party by checking whether the same package is called by other APKs, following similar methodology used by Oltrogge et al.~\cite{oltrogge2015pin} (described in more detail below). While there are a small number of public datasets of third-party library signatures, we find that these are generally too outdated or otherwise incomplete to be fit for purpose (e.g., LibRadar~\cite{ma2016libradar} was last updated in 2018).

We collect all packages containing a call to the Android Keystore key generation constructor \texttt{android.security.keystore.KeyGenParameterSpec.\\Builder(String keystoreAlias, int purposes)}. 
If a package is referenced by multiple APKs from different developers, we consider it to be third-party; otherwise, if it is referenced by only a single APK or by multiple APKs from the same developer, we classify it as a first-party package.

\vspace{1em}
\noindent\textbf{Obfuscation.} We observe a significant amount of obfuscation of package names where package names are shortened and anonymized (e.g., \texttt{o8} or \texttt{q1.x.a}), likely due to built-in obfuscation techniques available to developers in Android Studio and other widely used development tools.

Different packages may share the same obfuscated name, and so we exclude obfuscated packages from party analysis. To identify non-obfuscated package names, if a package name has at least one sub-component (i.e., character string separated by periods) of at least three characters in length, we consider it to be an authentic (non-obfuscated) package name.

\vspace{1em}
\noindent\textbf{Reachability.} Our call-graph generation methodology described above errs on the side of favoring false positives over false negatives (i.e., we would prefer to include a relevant API call that may be unreachable than to exclude a call that is used). To reduce the risk of false positives, once we have classified all packages as first-party or third-party, to determine whether a particular API call is reachable we trace backwards through the recorded call paths along the control flow. If there exists at least one path containing a call to first-party source code, we consider the API call to be reachable.

\subsection{Performance Measurements}
From the average developer's perspective, perhaps the most important consideration when choosing among different key storage APIs is performance. A natural corollary to surveying the usage of secure key storage is to investigate key storage runtime performance, particularly among different forms of hardware-backed key storage.

To conduct systematic performance measurements we wrote a benchmarking test application that performs symmetric and asymmetric key generation, message encryption, and message signing following canonical examples provided in Android documentation and Android's developer blog~\cite{android_keygenparameterspec,android_credentials_blog2013}. We use AWS Device Farm~\cite{aws_device_farm} to run our test application across a variety of Android devices.


\section{Secure Hardware Usage in Android}
\label{sec:secure_hardware_usage}

As the first step in our work, we conduct a comprehensive survey of all Android API calls relevant to key storage or trusted hardware, collecting arguments provided and relevant context (e.g., class and package name in which the call occurred). While we make every effort to retrieve the parameter argument via constant propagation in cases where static analysis initially returns the register value, this is not always possible and thus in the results below the parameter total for a particular API method call is generally lower than the method call total shown in Table~\ref{tab:keystore_api_refs}.

We further note that unless otherwise specified, statistics for API calls discussed throughout this section are not necessarily distinct: if a particular API call is located within a third-party library, this call configuration (e.g., parameters) is then duplicated in our findings for each call to this library (including across separate apps). We intentionally consider duplicates in our findings since our purpose is to understand the state of Android security and keystore usage in the wild, though for certain highly relevant calls we will distinguish between first-party (e.g., unique) calls and third-party library calls. Similarly, we will frequently distinguish between API usage as a percentage of total \textit{calls} for a particular API method and percentage of individual \textit{apps} containing at least one reference to the method in question since a single app can reference the same method numerous times.

\subsection{Overall Usage}
\label{sec:overall_usage}
Of the 486,234 in our dataset (apps currently in the Play Store with at least 10,000 downloads) which we were able to download and decompile, through keyword searching as described in~\S\ref{sec:keyword_filtering} we find 122,305 apps containing a reference to the Android Keystore API within their source code. This provides us with an upper bound of 25.15\% of apps within the Play Store using device trusted hardware. If we consider only the 159,241 apps self-reporting to the Play Store as collecting sensitive data, we find 69,583 apps referencing the Android Keystore API and the upper bound of trusted hardware use rises to 43.7\%.

In practice, these calls may be located within components of third-party libraries not referenced by the app, or within unreachable or legacy source code of the app itself. We then run our in-depth static analysis tool, KeyDroid, on all 122,305 apps flagged as directly referencing the Android Keystore API in some capacity to verify which calls are reachable and collect detailed statistics on how the API is used\footnote{A small number of APKs (2,365, or 0.48\% of our overall dataset) were flagged as containing the string ``AndroidKeystore'' but did not contain any references to the actual \texttt{android.security.keystore} API when searching the source code. After manual investigation we hypothesize that in most cases this is due to requesting the Android Keystore provider via a different Java API in potentially unreachable code (and so the Keystore API references along the call chain were removed at compilation). We include these APKs in our upper bound percentages reported above but exclude them from more in-depth analysis}. We are able to successfully analyze 116,555 apps, with the remaining 2.82\% erroring out for various miscellaneous reasons, most commonly exceeding the time limit.

The Android Keystore API further requires developers to specify an intended purpose at the time of key initialization and enforces this purpose when developers attempt to use the key (e.g., a key specified as being intended for encryption cannot later be used to sign). We find that of the 278,056 total init calls for which we were able to retrieve the purpose value, 92.31\% of keys are designated as being used for encryption and decryption only, while 5.60\% are used for signing or verifying message authentication codes.

A full list of all Keystore API endpoints and their total usage counts is shown in \S\ref{sec:api_totals}. We discuss most methods in more detail throughout this section. We further describe how usage varies by Play Store category in \S\ref{appendix:category_usage}, and describe alternative keystores used from a manual review of a subset of apps flagged as not using the Android Keystore API in~\S\ref{appendix:manual_analysis}.

\vspace{0.5em}
\noindent\textbf{StrongBox Usage.} We find that 22,875 of the 116,555 apps with any reference to the Android Keystore API (19.62\%) further contain a reference to the StrongBox API \texttt{setIsStrongBoxBacked\allowbreak(boolean)}, which is 4.7\% as a percentage of the overall dataset (and 5.03\% as a percentage of apps collecting sensitive data). However, since the API takes in a boolean parameter some of these instances may explicitly request \textit{not} to use StrongBox. To calculate how many apps \textit{enable} StrongBox, we are able to retrieve the argument value for 21,022 out of 24,630 calls and find that while 94.85\% of these instances request to use StrongBox, the remaining 5.15\% explicitly opt out of using StrongBox and storing cryptographic key material in the device’s secure element. Applying this percentage to the 22,875 apps referencing the API, we estimate that 22,367 apps, or 4.6\% of our overall dataset, request to use StrongBox for at least one key. This percentage rises slightly to 5.03\% if we consider only apps self-reporting collecting sensitive data. To better understand the context behind these choices without being hampered by source code obfuscation, we manually searched for instances of StrongBox disabling on GitHub~\cite{github_strongbox_disabled} as of January 2025. Of the 14 unique (i.e., non-fork) repositories which contained a call disabling StrongBox, two repositories included a comment citing performance reasons while 10 opted out without explanation. The Salesforce Android SDK, for instance, disables StrongBox as the runtime is "too slow" and therefore "not a good fit for [their] use case"~\cite{salesforce_strongbox,salesforce_strongbox2}. The remaining two instances disabled only if a \texttt{StrongBoxUnavailableException} was thrown and were therefore false positives.

The nested structure of Android key generation makes it difficult to reliably link a key generation call (which specifies the algorithm to be used) with the Android Keystore’s parameter specification using call objects, and simply checking whether both calls are located in the same method is too imprecise since a single method may generate multiple keys. Instead, we can indirectly estimate ciphers used for StrongBox specifically by linking key size with Strongbox usage. For the 98 keys which set both StrongBox and the key size and for which we are able to retrieve both parameter values, we find that 97 of 98 keys used StrongBox with an \texttt{AES-256} cipher while just one key used StrongBox to generate an \texttt{RSA-2048 key}, a distribution which again suggests runtime is a major consideration when using StrongBox.

\subsection{First-Party vs. Third-Party Usage}
Here we present a package-level analysis of the location context in which trusted hardware is referenced. In particular, we are interested in determining whether apps flagged as using trusted hardware are doing so as part of the core application source-code or because the hardware API is referenced indirectly as part of a third-party library. First-party usage indicates that developers have consciously chosen to store cryptographic key material in trusted hardware, while for certain third-party libraries (such as analytics libraries) developers may be unaware that this is occurring.

Overall, we find that the vast majority of Keystore API usages are located in third-party source code (definition provided in~\S\ref{interprocedural_call_graph_analysis}). Of a total of 199,156 calls to the Keystore init method located in non-obfuscated packages, we find that 94.69\% of calls originated in third-party libraries, while 5.31\% are located in first-party source code. This observed distribution is also true for SE usage. Of the 17,400 StrongBox calls located in non-obfuscated packages, 98.31\% are located in third-party libraries, while only 294 (1.69\%) are first-party calls within custom app source code.

\vspace{0.5em}
\noindent\textbf{Third-Party Libraries.} A natural follow-on question is \textit{which} third-party libraries referencing the trusted hardware API are most commonly used by apps. Table~\ref{tab:thirdparty_package_calls} in the Appendix shows the top 10 third-party libraries used by Android apps to reference the Keystore API. While several of the top 10 are security-focused libraries, four are primarily app development and analytics libraries, suggesting that the details of key generation and storage are abstracted from developers who may be unaware of what data is stored where.

\subsection{Key Authentication}
The Android Keystore API allows for a variety of authentication configurations to determine when a key can be accessed. The core authentication method \texttt{setUserAuthenticationRequired(boolean)} requires users to authenticate via any available form of device unlock (device pattern/PIN/password or biometric credentials) for any cryptographic operations using a private key~\cite{android_setuserauthrequired}. More specialized API methods allow developers to require a specific form of authentication (e.g., biometric authentication only) and to set the duration during which the authentication is valid.

We find that 15.84\% of keys stored in the Android Keystore require some form of user authentication prior to granting access, with 2.78\% requiring biometric authentication specifically (and disallowing any other form of authentication, such as device passcode).

By default, if a key requires any form of authentication then a user must authenticate each time the key is used. To provide a more user-friendly configuration, the Keystore API allows developers to set a validity duration period in seconds during which the key can be reused without any need to reauthenticate.
21.75\% of keys require the user to authenticate each time they initiate an operation requiring key access, the most secure configuration but also one that can use friction for the user experience. For calls that set a specific duration, the most popular durations were 5 seconds (set by 38.53\% of keys which set a duration) and 1 hour (set by 4.45\% of keys). A significant percentage of API calls set very short validity durations: 13.2\% of calls that set a duration set it to 3 seconds or less, meaning that the user can only reuse the key within the next few seconds. For some use cases, unless the user proceeds very quickly this is effectively the same as requiring authentication each time.

As an alternative to requiring a user to provide information to authenticate, a user can instead approve a pop-up message via the \texttt{setUserConfirmationRequired(boolean)} API before proceeding. As a standalone API this does not require the individual approving the message to provide any information indicating that they are indeed the device owner (i.e., they need only tap to approve), but it can be used in combination with the authentication APIs described above to provide cryptographic certification that a user has approved a certain action. However, we find that very little use of this feature: of the 26 calls to the \texttt{setUserConfirmationRequired} API detected where we were able to retrieve the argument value, only two of them enabled confirmation (with the remaining 24 disabling).


\subsection{Implementation Security}


\noindent\textbf{Ciphers.} Of 232,283 key generation calls to Android cryptographic APIs requesting the Android Keystore as provider, 63.51\% requested an AES key, 34.48\% requested an RSA key pair, and 0.9\% of keys requested an EC key pair. Table~\ref{tab:keystore_ciphers} in the Appendix shows the full list of requested ciphers and their respective usage counts. 

As a point of comparison, of the 20,042 calls requesting the AndroidOpenSSL software-backed provider, 99.74\% generated an RSA key pair with just 51 generating an AES key. We hypothesize that developers avoid hardware-backed key storage for asymmetric encryption  out of performance concerns, which we discuss further in subsequent sections.

The Android Keystore API also includes legacy ciphers for backwards compatibility and interoperability, some of which have since been deprecated. 3DES, for instance, was simultaneously added and deprecated in API level 28~\cite{keystore_3DES}. In our analysis, we fortunately find very few instances of apps using insecure or legacy ciphers. In particular, we find no instances of \texttt{3DES} or \texttt{HMAC-SHA1} even though these ciphers are available within the Android Keystore~\cite{keystore_keyproperties}.

\vspace{0.5em}
\noindent\textbf{Defaults.} Android Keystore API defaults are significantly more secure than those of software-based Java cryptography APIs historically available in Android. For instance, if a developer requests an AES cipher without specifying the accompanying encryption mode(s) as in \texttt{javax.crypto.Cipher.getInstance(``AES'')}, Java's Cipher API defaults to AES with ECB mode, an insufficiently random configuration~\cite{android_ecb_default}.

Android Keystore, on the other hand, disallows various insecure cryptographic operations by default, including using ECB mode in symmetric encryption, RSA encryption/decryption without proper padding, and using an insufficiently random IV~\cite{android_setrandomizedencryptionrequired}. All of the six essential rules in cryptography laid out by Egele et al.~\cite{egele2013empirical} (e.g., do not use ECB mode with symmetric encryption, do not use a non-random IV for CBC) in 2013 are not possible within the Keystore API by default. Unless the developer explicitly disallows randomized encryption, many of the same configurations that run smoothly or are even the default in Java's software APIs will throw an \texttt{InvalidKeyException} with the Android Keystore. In addition to disallowing insecure configurations by default, the Android Keystore API is designed such that it requires developers to provide specific configurations instead of providing only a high level cipher (e.g., for symmetric encryption a developer must specify the block mode(s) and encryption padding at the point of key generation using the designated \texttt{setBlockModes} and \texttt{setEncryptionPaddings} APIS~\cite{android_setblockmodes,android_setencryptionpaddings}). Android Keystore then verifies that the configuration provided is valid, sufficiently secure, and compatible with the specified key purpose.

\vspace{0.5em}
\noindent\textbf{Randomized Encryption.} 
It is possible, however, for developers to circumvent Android Keystore's secure default settings and implement known insecure configurations by setting Keystore's \texttt{setRandomizedEncryption(boolean)} API~\cite{android_setrandomizedencryptionrequired}, which mandates configurations must be sufficiently randomized to provide indistinguishability between ciphertexts given chosen plaintexts (e.g., \textit{IND-CPA}), to false. In general, disabling this API means that the same plaintext encrypted with the generated key may produce similar or identical ciphertexts.

\begin{figure}[t]
   \centering
    \includegraphics[width=1\columnwidth]{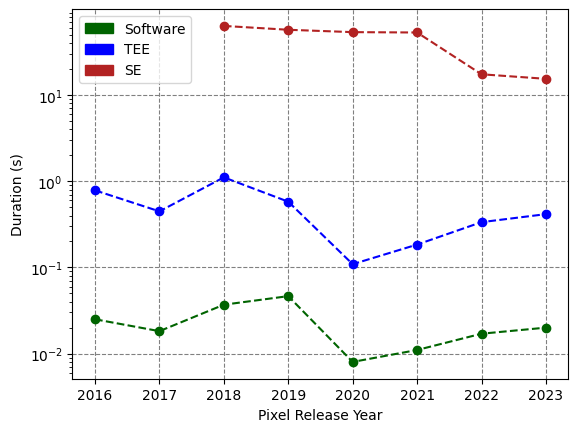}
    \caption{Performance evolution of encrypting 1 MiB with AES-GCM in Pixel devices. Each data point corresponds to the Pixel device released in that year (e.g., 2023 represents measurements taken from the Pixel 8). The y-axis is log-scaled.}
    \label{fig:pixel_performance_evolution}
\end{figure}

\begin{figure}
    \centering
    \includegraphics[width=1\columnwidth]{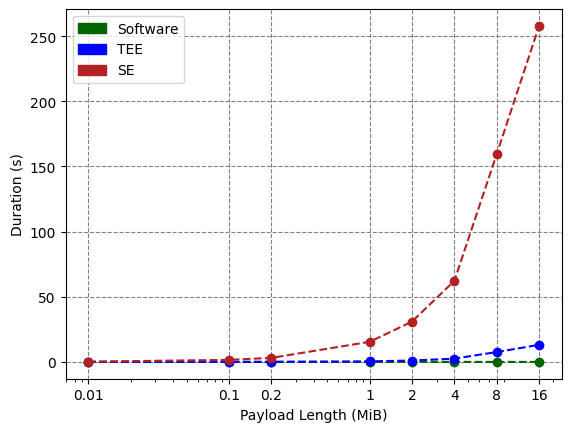}
    \caption{Execution times of AES-GCM-256 encryption as a function of message length on the Pixel 8. The x-axis is log-scaled. The precise numerical runtimes are shown in Table~\ref{tab:encryption_length_runtime} in the Appendix.}
    \label{fig:message_length_encrypt_log}
\end{figure}

We find that 77.94\% of calls to the randomized encryption API disable the setting (a relatively unsurprising result given that it is enabled by default, and so referencing the API with \texttt{True} as the argument has no effect). When estimating how this configuration is distributed as a percentage of all hardware-backed keys, however, given that there were 30,245 references to the randomized encryption API endpoint we estimate that approximately 8.45\% of all Android Keystore-backed keys disable \textit{IND-CPA}, a surprisingly high percentage given that this configuration violates a core cryptographic security property.

There are a handful of scenarios in which a developer may deem it necessary to disable this requirement (for instance, if a custom IV is needed), though the API documentation suggests alternative workarounds to avoid disabling randomized encryption for several common cases~\cite{android_setrandomizedencryptionrequired}. To investigate this further, we identify the ten most-used libraries containing a call disabling randomized encryption and manually review each, though we find only three are public without significant reverse engineering: (1) \texttt{com.ama\-zonaws.internal.keyvaluestore}, AWS's internal keystore which generates a secure key configuration (\texttt{AES/GCM/NoPadding}) but disables randomized encryption because the API ``does not work consistently in API levels 23-28''~\cite{aws_randomizedenc}, (2) \texttt{com.apptentive.android.sdk.encryption\-.resolvers}, a customer engagement platform which uses a custom initial vector (IV) and thus is required to disable randomized encryption~\cite{apptentive_randomizedenc}, and (3) \texttt{dev.mcodex.RNSensitiveInfo}, a React Native wrapper library which disables randomized encryption for a basic AES/GCM/NoPadding configuration as AWS did~\cite{react_randomizedenc}. Our results are inconclusive as we manually searched Android bug trackers for historical issues with randomized encryption API and could not find any relevant results, but these reported issues with consistency may be an area for the Android team to issue public guidance.


\vspace{0.5em}
\noindent\textbf{Key Attestation.} Android Keystore allows developers to require key attestation, which verifies that keys are indeed stored in device hardware~\cite{android_setkeyattestation}. We find 2,724 calls to \texttt{setAttestationChallenge(byte[])}, indicating that 0.98\% of all keys generate an attestation certificate chain. While still a relatively small percentage, this nonetheless represents a significant increase from Imran et al.~\cite{imran2022sara} who previously scanned a randomly sampled subset of 112,886 Android apps for attestation in January 2021 and found only 5 apps using key attestation.

\section{Key Storage Performance}
Having surveyed the current usage of trusted hardware, in order to judge when hardware-backed key storage \textit{should} be used we must first understand what performance differences, if any, exist compared with software-backed key storage. Unfortunately, to the best of our knowledge Android does not currently publish any empirical statistics evaluating the performance of software or hardware keystores.

To conduct our own measurements, we use AWS Device Farm~\cite{aws_device_farm} to measure the runtime performance of key storage options across a variety of Android devices. Our test app calculates the runtime performance of each individual operation for the following three keystores: the device's default software-based keystore (Bouncy Castle for the Pixel XL and AndroidOpenSSL for all other devices), the Android Keystore using the default TEE configuration, and the Android Keystore using a SE (StrongBox Keymaster). The numbers reported below for each operation represent the average performance across 100 distinct iterations.

\subsection{Performance Evolution}
We first measure how key generation and encryption performance has changed over time using Google’s flagship Pixel device line from 2016 through 2023.

\vspace{1em}
\noindent\textbf{Key Generation.}
Our results show that symmetric key generation has a negligible performance impact regardless of the keystore used. In the most recently released Pixel device, the Pixel 8, generating an AES-GCM-256 key takes $0.002 s$ in Android’s software keystore, $0.021 s$ in Android’s TEE keystore, and $0.071 s$ in Android’s SE keystore, StrongBox. We observe similar runtimes for older Pixel devices. While this represents a large percentage difference, the real runtime impact is negligible given the small execution times.
Runtime differences are more significant with asymmetric encryption: in the Pixel 8, generating an RSA-2048 key takes $0.21 s$ in a software keystore, $1.93 s$ in Android’s TEE keystore, and $9.22 s$ in StrongBox.

\vspace{1em}
\noindent\textbf{Key Encryption.}
Figure~\ref{fig:pixel_performance_evolution} shows the comparative performance of encrypting a randomly generated 1MiB payload with AES-GCM-256 with no padding across Pixel devices released between 2016 and 2023. Android introduced a secure processor beginning with the Pixel 3, and consequently StrongBox measurements are only shown from 2018 on.

The performance impact of software-backed encryption and TEE-backed encryption has roughly stayed the same over time, with the original Pixel and the most recent Pixel 8 reporting TEE measurements of 0.78 and 0.41 seconds respectively. For a payload of 1MiB or smaller there is a negligible difference between running cryptographic operations inside a TEE and running them natively in terms of what is observable to the end user, which has been the case since the initial release of the Android Keystore API. 

StrongBox encryption, however, is significantly slower than the other two keystore types. In the Pixel 8, for a 1 MiB payload StrongBox symmetric encryption takes an average of 15.43 s while TEE encryption takes 0.42 s and encryption using a software-backed key takes just 0.02 s. StrongBox performance has improved over time, and so execution times are even longer in older devices: the initial Pixel 3 (released in 2018) has a symmetric encryption runtime of $63.43 s$ which held reasonably steady until the release of the Pixel 7 in 2022 where the performance dropped significantly to $17.42 s$. The sharp performance improvement between the Pixel 6 and Pixel 7 is somewhat surprising since both devices use Google's in-house Titan M2 security chip~\cite{android_pixel6}. The Pixel's main processor changed from Google Tensor in the Pixel 6 to Google Tensor G2 between the 6 and 7 devices, however, and it is possible that the main Tensor G2 processor is able to communicate with the Titan M2 chip more efficiently.

For asymmetric encryption, we measure Pixel 8 performance across keystores on a very small payload of 256 bits (i.e., the use case where a software-backed AES key is encrypted by a hardware-backed RSA key). We find that asymmetric encryption incurs very little performance overhead on minuscule payloads regardless of keystore, with TEE encryption taking an average of 0.0065$s$ and StrongBox encryption taking 0.0125$s$ on average.

Overall, symmmetric encryption using a SE-backed key is roughly 35 to 55 times slower than encryption using a TEE-backed key depending on the device, likely due to the cost of round-trip communications between the main processor and secure processor. This finding somewhat contradicts Android’s official documentation, which qualitatively describes StrongBox as ``a little slower'' as previously mentioned in~\S\ref{sec:secure_element}. On the most recently released Pixel device, however, basic symmetric encryption of a 1MiB payload within StrongBox takes around 37 times (and 15 seconds) longer than the same operation within a TEE.

\subsection{Performance vs. Payload Length}
We further measure the impact of message length on encryption performance. Figure~\ref{fig:message_length_encrypt_log} shows the performance of payload sizes between 1MiB and 16MiB for the Pixel 8 (again using AES-GCM-256 with no padding). In this experiment we used the average of 10 iterations for payloads 4MiB and above (instead of 100 iterations as with other experiments) due to rapidly increasing execution times.

While encryption runtime increases linearly with message length for all three keystore types, StrongBox runtime quickly becomes unmanageable for large lengths. A relatively small payload of 0.1 MiB takes the Android Keystore $0.08 s$ to encrypt using the TEE and takes StrongBox $1.59 s$. A 4MiB payload, however, will take StrongBox roughly 1 minute to encrypt, while the TEE-backed keystore can encrypt the same payload in just $2.56 s$, making the TEE viable even for larger message lengths. A software-backed keystore provides the best performance by far as expected, encrypting payloads of up to 16 MiB in just 0.3 seconds given that all operations are in-process with no IPC calls or context switch. Table~\ref{tab:encryption_length_runtime} in the Appendix contains the TEE and SE execution times and standard deviations for all message sizes tested on the Pixel 8 (shown visually in Figure~\ref{fig:message_length_encrypt_log}).

Execution times for message signing are similarly cost-prohibitive using StrongBox. As shown in Table ~\ref{tab:signing_length_runtime} in the Appendix, while StrongBox needs only 1 second to sign a small payload of 0.1 MiB, this runtime increases to 9 seconds for a payload of 1 MiB and 35.91 seconds for a 4 MiB payload. In contrast, a TEE is able to sign a 4 MiB message in 1.76 seconds, making it roughly \texttt{20x} faster than StrongBox.

\begin{figure}
    \centering
    \includegraphics[width=1\columnwidth]{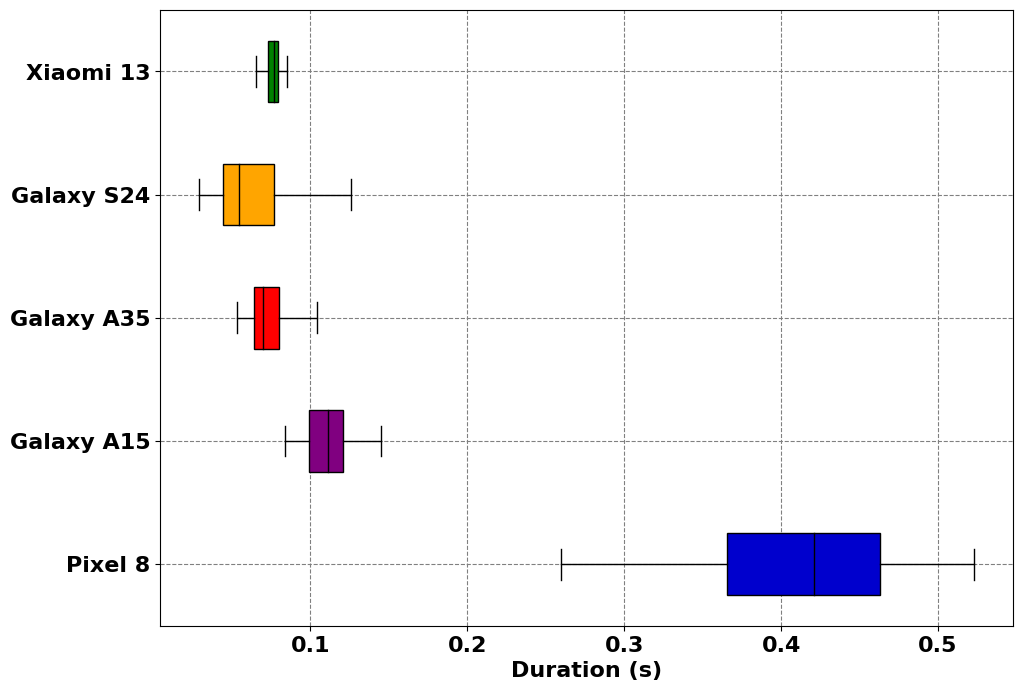}
    \caption{Runtime duration of encrypting 1 MiB with AES-GCM within a TEE across a range of Android devices recently released in the past two years. While four of the five devices cluster around 0.1 seconds, the runtime of the Pixel 8 is noticeably longer and with a wider range.}
    \label{fig:tee_performance_all}
\end{figure}

\subsection{Cross-Provider Performance}
We further investigate how Pixel performance compares with other commonly used mobile devices in the Android ecosystem. Figure~\ref{fig:tee_performance_all} shows TEE performance for symmetric encryption across a range of Android devices, including Samsung and Xiaomi. The five devices measured were chosen by selecting the most recently released device across all device lines available through AWS Device Farm. Four of the five devices measured (Samsung Galaxy A15, Samsung Galaxy A35, Samsung Galaxy S24, and Xiaomi 13) consistently report runtimes around 0.1 seconds for TEE-backed symmetric encryption, while the Pixel 8’s average runtime is 0.41 seconds.

While the Galaxy A15, Galaxy A35, and Xiaomi 13 devices do not include a secure element~\footnote{Samsung first introduced a secure processor in 2020 but only within its Galaxy S series~\cite{samsung_galaxys20}. Devices recently released as part of other series (such as the Galaxy A15 and Galaxy A35 devices, introduced in December 2023 and March 2024, respectively) do not include a secure processor (and thus throw a \texttt{StrongBoxUnavailableException} if a developer attempts requests to store keys in the StrongBox). We confirmed this through our own tests.}, we compare StrongBox performance between the Pixel 8 and the Samsung Galaxy S24 (released in January 2024) and find a noticeable difference in performance in symmetric encryption. As previously discussed above the Pixel 8 takes $15.43 s$ to execute AES-GCM for a 1MiB payload, while the Galaxy S24 takes $26.39 s$, or close to twice as long. Curiously, the inverse is true for these two devices when considering TEE performance as shown in Figure~\ref{fig:tee_performance_all}: the Pixel 8 takes $0.41 s$ to execute symmetric encryption using a TEE-backed key, while the Galaxy S24 takes far less time at $0.06 s$, illustrating the nuances and complexities of each individual device’s processor(s) and other hardware.

\section{Developer Survey}
\label{sec:survey_methodology}
To better understand why Android developers opt not to use hardware APIs, we conducted a large-scale developer survey in August 2024 for apps flagged as matching either of two trusted hardware configurations of interest. This study was approved by our department's ethics committee (equivalent to IRB), and all response data was aggregated and anonymized (see \S\ref{sec:ethics} for an in-depth ethics discussion). The survey questions are given in \S\ref{sec:appendix_survey}.

We are interested in two broad categories of apps and conducted separate surveys for each: (1) \texttt{Sensitive-NonKeystore}: apps that self-reported as collecting sensitive user data but did not use Android's trusted hardware APIs (either in first-party \textit{or} third-party components) and (2) \texttt{StrongBox-Disabled}: apps that referenced the Android Keystore API in a first-party context but explicitly disabled StrongBox for at least one key (e.g., they requested to only use TEE-backed key storage). Both of these high-level configurations indicated that the app developers may have made a conscious decision not to use some form of trusted hardware.

For the first (\texttt{Sensitive-NonKeystore}) configuration, we surveyed a random sample of 10,000 developers via email using the contact information given on the Play Store, and have received $n = 42$ responses at the time of writing. We attribute the low response rate in large part to the use of Play Store app support email addresses, which are often read by a customer service team (if one exists) and who may not pass on our survey request to developers. 

Of the 42 responses, 18 respondents reported that one factor in opting not to use trusted hardware APIs is that their app does not store credentials and/or deemed the security benefits unnecessary given the type of user data collected. Three respondents reported general performance concerns, while 14 respondents indicated that legacy development or compatibility reasons were prohibitive factors, reporting either a desire to maximize devices the app can run on or that the app was developed prior to the Android Keystore API release date in 2013. One such developer specified that their app uses SQLite due to ``lack of knowledge [of the Android Keystore API] at the time of development and difficulties for migrating later.'' Notably, API usability did not appear to be a widespread concern---just two of the 42 respondents indicated they had found the Keystore API difficult to use.

For the \texttt{StrongBox-Disabled} configuration, after filtering out third-party StrongBox calls we identified $n = 25$ apps matching a StrongBox-disabled configuration. Unfortunately we received no responses for our \texttt{StrongBox-Disabled} survey, a relatively unsurprising response rate given our restriction of the dataset to first-party disabled calls limited our sample size. Even so, our manual review of disabled StrongBox configurations on GitHub described in~\S\ref{sec:overall_usage} has also provided a window into developers' thought processes.

\section{Limitations}
\label{sec:limitations}
Here we acknowledge the following limitations of our analysis and describe steps taken to mitigate these limitations.

\vspace{0.5em}
\noindent\textbf{Accuracy of static analysis:} As with prior work in Android app analysis, our research is subject to the inherent technical limitations of static analysis. Given that we only have access to packaged bytecode instead of the original source code, we cannot guarantee that certain source code components have not been obfuscated, though it is unlikely that this would be the case for Android system APIs. Static analysis cannot reliably detect whether a particular component is executed at runtime (i.e., dead or legacy code), but this is a natural trade-off with the scale of our work. Modern compilers and widely used app optimization tools are highly effective at removing unused source code and so we anticipate app bytecode is unlikely to contain unreachable code at the point of our analysis.
Dynamic analysis would further preclude studying certain categories of apps, such as financial apps, since we cannot create test financial accounts for regulatory reasons.

\vspace{0.5em}
\noindent\textbf{Necessity of high-level analysis:} The scale of our work (downloading and analyzing around half a million apps) necessarily means that our analysis will be comparatively high-level. In particular, static analysis is unable to automatically detect the semantic application context in which a trusted hardware API call occurs, including what particular data is being stored within the hardware element and how keys generated are being used, or to guarantee that the flagged API call is used to protect sensitive data at runtime (e.g., an app might import a marketing analytics API that in turn references the Keystore API for processing analytics data). However, in our work we are primarily interested in which apps choose \textit{not} to use trusted hardware, particularly SEs, and why. Our results provide an empirical upper bound on secure key storage usage and provide comprehensive data on API usage and performance across the Android ecosystem as a whole.


\section{Discussion}
\label{sec:discussion}

\noindent\textbf{Trusted hardware usage is still comparatively low}: 
While both industry and government have launched various initiatives encouraging developers to move towards trusted hardware~\cite{cisa_secure2023,android_ready_se,ukgov_code_practice}, usage remains stubbornly low even among apps that, by their own admission, collect potentially sensitive user data. 
Just 43.7\% of apps processing sensitive data use any form of trusted hardware, and almost all of this usage comes from third-party components. While some of these apps may be collecting relatively benign data (such as a user's name) or may rely primarily on a remote server to handle most cryptographic operations instead of storing data on device, this is still a comparatively low rate given there is little to no performance drawback for common cryptographic use cases in a TEE-backed keystore.

Additionally, the vast majority of apps using hardware-backed storage use a TEE instead of an SE (43.7\% compared to 5.03\%). Put another way, while Google's public goal is to make the SE the "lowest common denominator" in credential storage~\cite{android_ready_se}, as of 2024 we observe that only around 10\% of apps using trusted hardware at all are using the SE at least once. As side-channel attacks become ever more sophisticated and effective~\cite{bursztein2023generic}, it is even more important for applications to use the most advanced storage available to protect data.

\vspace{0.5em}
\noindent\textbf{Android Keystore API provides more secure defaults:}
In addition to the protection secure hardware provides against OS compromise, the Android Keystore API also offers significantly more secure defaults than similar Java cryptographic APIs. Android Keystore mandates an \texttt{IND-CPA}-secure configuration by default, disallowing insecure configurations that have plagued other cryptographic APIs~\cite{egele2013empirical, focardi2018mind}. Android also runs checks to ensure the security and validity of a configuration, including cross-referencing the stated purpose of a key with which it is generated (e.g., EC keys cannot be used for encryption and decryption, only signing). While it is still possible for developers to circumvent this default (as 8.45\% of them do), this nonetheless requires a conscious decision by the developer. Android Keystore's default settings alone make it a security improvement over other cryptographic APIs.

\vspace{0.5em}
\noindent\textbf{TEE-backed storage performance is viable for small-to-medium message sizes}: We find a negligible difference (<0.5 seconds) between TEE-backed and software-backed cryptographic operations for payloads less than 1MiB, empirically confirming that in common scenarios hardware key storage runtime is not a prohibitive factor when using the Android Keystore API. A TEE keystore can thus provide significant security benefits with minimal performance impact, providing an ideal trade-off between enhanced security and performance overhead for most app use cases. 

\vspace{0.5em}
\noindent\textbf{Need for public performance evaluations of StrongBox:} In comparison to the TEE, Android’s SE demonstrates significantly worse processing time for all but the smallest payloads. If we consider acceptable processing times to be less than three seconds, StrongBox can only encrypt message sizes of roughly 0.2 MiB or less even in the most recently released Pixel devices. For comparison, a TEE can encrypt message sizes of up to around 2 MiB within the same time frame. Our performance measurements, static analysis of symmetric versus asymmetric usage patterns, and manual review of calls disabling StrongBox all strongly suggest that performance is a prohibitive factor in using SEs in practice. 5.15\% of developers referencing the Android Keystore API explicitly opt out of using StrongBox (as in the Salesforce example in~\S\ref{sec:overall_usage}).

Even so, StrongBox's execution time may be entirely reasonable in cases with very small payloads: for instance, an app may use StrongBox to encrypt a different cryptographic key. Equally, developers may evaluate overhead cost differently depending on whether it is a one-time operation (e.g., initial login) or a repeated process. Developers need quantitative information in order to make case-by-case decisions, a gap which our work fills. Most importantly, Android's documentation arguably understates the depth of the performance drawbacks of SEs, making it more challenging for developers to make an informed decision. Updated, empirical performance measurements based on contemporary device measurements should be publicly available and easily accessible to developers in place of the ambiguous language currently used in the documentation., which may also have led developers to opt out of using StrongBox as a precautionary measure.
\section{Related Work}
\label{sec:related}

\noindent\textbf{Android App Analysis:} Most prior work studying security and privacy in Android apps has used metrics such as permissions requested~\cite{felt2011android,li2021android,reardon201950} and traffic analysis~\cite{fahl2012eve,possemato2020towards,oltrogge2015pin} and has often overlooked data storage, even when investigating overall app security~\cite{kollnig2021iphones}. For instance, Gilsenan et al.~\cite{gilsenan2023security} studied security issues in two-factor authentication (2FA) apps and recommended that apps use the Android Keystore, but did not investigate how apps actually do store their keys.

Egele et al.~\cite{egele2013empirical} studied cryptographic misuses in Android applications in 2014, but looked only at the software-backed Java Cryptographic Architecture APIs (presumably due to the timing of the work, since the initial Android trusted hardware API was only released in 2013). They noted at the time that both Java and Android JCA APIs allowed a developer to specify only the encryption algorithm (e.g., AES), in which case Java and Android used ECB mode with PKCS7Padding as the default. 
Focardi et al.~\cite{focardi2018mind} similarly analyzed the confidentiality and integrity properties provided by various software-backed Java keystores in 2018.

There have been a handful of studies focusing on particular subsets of hardware-related API usage in Android. Bianchi et al.~\cite{bianchi2018broken} conducted an empirical survey of Android’s Fingerprint API and found very low adoption rates, with just 424 of 30,459 popular apps scanned using the API. Imran et al.~\cite{imran2022sara} ran a keyword search for the key attestation API on a subset of apps in sensitive categories (e.g., finance, communication, medical), finding that of 112,886 apps only five use key attestation. Concurrently to our work, Bove~\cite{bove2023large} conducted a high-level study on various TEE-based Android APIs (including the Biometrics and Digital Rights Management APIs) on a randomly sampled subset of Play Store apps and found that 32.0\% of apps analyzed contained a call to the Keystore API (excluding gaming apps), but only measured the binary question of whether an app contained any Keystore API call without investigating usage specifics. Additionally, a particular focus of our work is comparing TEE and SE APIs in both usage and performance.

Coojimans et al.~\cite{cooijmans2014analysis} systematized high-level security properties of Android key storage options in 2014, observing that while Android’s TEE-backed key storage provides device binding (i.e. prevents keys from being extracted from the device) where software keystores are vulnerable, the implementation of the TEE keystore made it possible for an attacker with root permissions to use other apps’ keys (i.e. did not effectively provide app-binding). Our work expands on this discussion to consider new forms of hardware (namely, SEs) that were not available when Coojimans et al. surveyed Android key storage in 2014.

\vspace{0.5em}
\noindent\textbf{Trusted Hardware Performance:} To the best of our knowledge, Android does not provide official quantitative assessments of trusted hardware performance. There has been a small amount of prior work measuring specific aspects of trusted hardware performance in Android as supporting experiments demonstrating the viability of a proposed cryptographic scheme. Hugenroth et al.~\cite{hugenroth2023sloth} measured the performance of HMAC execution in SEs on Android and iPhone devices to confirm their proposed key stretching scheme was feasible on contemporary devices, observing that time elapsed increases linearly with input length and that a 10 KiB payload takes approximately 1 second to execute in the Pixel 3 SE. In our work, we present the first comprehensive, longitudinal analysis of the performance of various key storage schemes, measuring the comparative performance of all widely used ciphers across the three major key storage options for developers (a software-backed Java Keystore, Android’s TEE Keystore, and Android’s StrongBox SE API).

\section{Conclusion}
\label{sec:conclusion}

This work presents the first comprehensive, large-scale survey of trusted hardware usage and performance in Android devices. While even the most secure trusted hardware configuration is ultimately best-effort as developers have to contend with available device hardware, we find that a significant percentage of apps, including those self-reporting to the Play Store as collecting sensitive user data, do not make use of the Android Keystore trusted hardware API. Our performance results show that TEE-backed key storage is viable for all but very large payloads, removing one of the most significant barriers to adoption. Our results provide app developers with concrete performance data to encourage adoption and ultimately to enable them to make an informed decision for their individual use case(s).

\section{Open Science}
\label{sec:dataset}
In compliance with the open science policy and in the interest of open access, we have open sourced all data used in our analysis, including our APK dataset, keyword search and call graph analysis results files for all individual APKs, and all source code and testing scripts used. Our performance benchmarking test app and all runtime logs are also publicly released. Our dataset can be accessed here: <redacted for review>.

\section{Ethics Considerations}
\label{sec:ethics}

We carefully considered the ethics of all components of our research. All static analysis is conducted on publicly available data, including both published apps and Play Store metadata. We intentionally keep our discussion of usage analysis aggregated and comparatively high-level to avoid the perception of targeting specific apps for potentially insecure configurations, though we open-source all analysis results as described in~\S\ref{sec:dataset}.

\vspace{0.5em}
\noindent\textbf{Developer survey.} As part of our developer survey, we sent a single initial email to each app developer in our random sample of 10,000 apps, after filtering the random sample to ensure that we only selected one app from each developer (i.e., a developer would not receive more than one email). All email addresses were retrieved by scraping the Play Store and were intentionally provided as a point of contact for the public. We did not send follow-up or reminder emails to avoid spam.

Our email clearly identified ourselves as academic researchers, including institutional affiliation, in the first sentence and emphasized that we were conducting a voluntary research study in which all responses were anonymous. If recipients clicked on the survey link, we further included an informed consent statement at the beginning of the survey that included similar information in greater detail, and invited respondents to reach out directly over email with any questions.

Our developer survey was approved by our computer science department's ethics committee (the effective equivalent of an Institutional Review Board) after the committee reviewed the proposed survey questions, email, and informed consent statement appearing at the beginning of the survey. Additional specifics of the methodology are described in~\S\ref{sec:survey_methodology}. We opted not to financially compensate participants in order to adequately preserve anonymity in line with similar work involving large-scale surveying of Play Store developers but took care to keep the survey length to a minimum (estimated 2-3 minutes) to be respectful of developers' time.

\section*{Acknowledgments}
Jenny Blessing is funded by Entrust and Nokia Bell Labs.
Ross Anderson made important contributions to the ideas contained in this paper. 
Unfortunately he died on 28th March 2024 before the final version was written; any errors remain our own.

\bibliographystyle{plain}
\bibliography{references}

\appendix
\clearpage
\section{Appendix}

\subsection{Trusted Hardware Best Practices}
\label{appendix:best_practices}
\noindent\textbf{Legal Mandates:} In certain industries, developers have to abide by a heavy patchwork of regulatory standards governing data collection and processing. In the U.S. (the region in which our application dataset and ranking information are collected) since 1996 the medical sector has been governed by the Health Insurance Portability and Accountability Act (HIPAA)’s Security Rule~\cite{hipaa_security_rule}, which specifies minimum security standards that health service providers must meet. The financial sector has a variety of SEC regulations and longstanding laws they must comply with, such as the global Payment Card Industry Data Security Standard (PCI DSS) that governs processing and storage of credit card data.

Regulatory standards in the financial and medical industries generally require that data is encrypted at rest. Specific implementations, such as use of secure hardware to store credentials, are usually not mandated directly. For instance, the American Medical Association acknowledges that since security is an ``evolving target, and so HIPAA's security requirements are not linked to specific technologies or products''~\cite{ama_hipaa}. Rather, regulation often encourages adoption indirectly, such as a law that mandates a security standard only provided by the hardware-backed storage mechanism. For instance, an industry may be required to use a FIPS-compliant random number generator~\cite{hipaa_fips}, which on a particular mobile device is only available via the HSM API. Moreover, even in cases where regulations provide little specific guidance, providers operating in heavily-regulated sectors are generally motivated to prioritize security within their product to keep pace with the sector in which they operate and minimize the risk that they could be charged with running afoul of the law.

\medskip
\noindent\textbf{Developer Guidelines:} Android’s published security guidelines for developers~\cite{android_security_guidelines} recommends developers use the Android Keystore for long-term or multi-use keys, and the OWASP Mobile Application Security Testing Guide~\cite{owasp_guide} recommends that developers “should always rely on” available secure hardware to store and use encryption keys. To audit app security, Android’s ``app security improvement program''~\cite{app_security_improvement} further scans all applications for various potential security issues upon initial submission and subsequent updates, including well-known vulnerabilities (e.g., Logjam), unsafe encryption modes, and insecure connection issues. We are not aware of any analysis of key storage.

\subsection{Play Store Data Safety Labels}
\label{appendix:data_safety}
In this paper we use the Play Store's data safety label information to determine which apps process sensitive data, and therefore which apps may be expected to make use of secure key storage.

\vspace{1em}
\noindent\textbf{Published data safety information.} 
According to Google’s developer documentation, ``all developers that have an app published on Google Play must complete the data safety form'' (including apps that self-report not collecting user data)~\cite{google_dev_data_safety}. In practice, we find that only 74.47\% (342,872/460,362) of apps have submitted a data safety form at the time of scraping in March through April 2024\footnote{The slight difference in number of apps for which we attempted to retrieve a data safety label (460,263) vs. number of apps downloaded and decompiled (486,234) is due to apps that were available in the Play Store at the time we began scraping apps themselves in October 2023 but had been removed by the time we began scraping app data safety pages in March 2024.}.
Khandelwal et al.~\cite{khandelwal2023unpacking} had previously conducted a large-scale analysis of Play Store data safety labels in May 2023 (approximately one year earlier) and found that only 46.8\% of apps reported any data, so we note the percentage of apps providing a data safety label has increased significantly from approximately a year earlier, though it is still noticeably far from satisfying the Play Store mandate.

For apps that have data safety information, we classify each app as ``sensitive'' or ``benign'' based on the types of data the developer has reported. 
Google uses 14 high-level data type categories, such as location, financial information, audio files, etc.~\cite{google_dev_data_safety} We consider an app to be sensitive if it collects any information from 12 of these 14 data types. We exclude the final two categories, “App info and performance” (defined by Google as crash logs and other app performance data) and “Device or other IDs” (e.g. MAC address or Firebase ID), since we are interested in whether developers are intentionally collecting sensitive user data relating to specific individuals, which we broadly define as user-provided data.
Based on this classification, we find that of the 342,872 apps reporting data safety information, 46.75\% are sensitive (and therefore 53.25\% are benign).

\vspace{1em}
\noindent\textbf{Developer self-reporting.}
Google uses a somewhat counter-intuitive notion of what constitutes data collection: instructions to developers state data is considered to be collected if it is transmitted ``from your app off a user’s device''~\cite{google_dev_data_safety}, and user data that is only processed and stored locally does not need to be reported as ``collected''.
In short, it is possible that an app that processes sensitive user data locally (and may therefore be expected to use some form of hardware-backed key storage) yet this app would not be listed as collecting sensitive data.

Therefore, by using the information in the data safety labels there is a risk our analysis excludes apps which do in fact process sensitive data. 
Nevertheless, we argue that there is much to be gained from understanding how the Keystore API is used by those apps which state they process sensitive data: if these apps do not make use of hardware-based secure storage, we hypothesize that it is unlikely that those apps which do process sensitive data, but do not declare it in their data safety label, process such data securely.

There are both benign and malicious reasons for developers inaccurately reporting their use of sensitive data. 
For example, developers may be unaware of the data collected by third-party libraries or wish to avoid highlighting the data their app collects in their submission, and thus may understate data collected. Conversely, it is also possible that developers may err on the side of \textit{overstating} the sensitivity of the data they collect to ensure they are in compliance with Play Store policies.
In principle, Google can often verify whether an app collects sensitive data (or not) and spot any differences between app behavior and reported collection.
If discrepancies are found, Google has the ability to block app updates or remove the app from the Play Store altogether.
We are unable to determine the extent to which such verification and enforcement takes place and therefore validate the correctness (or otherwise) of the data safety label information.

\subsection{Usage by Category}
\label{appendix:category_usage}
\begin{figure}
    \centering
    \includegraphics[width=0.95\columnwidth]{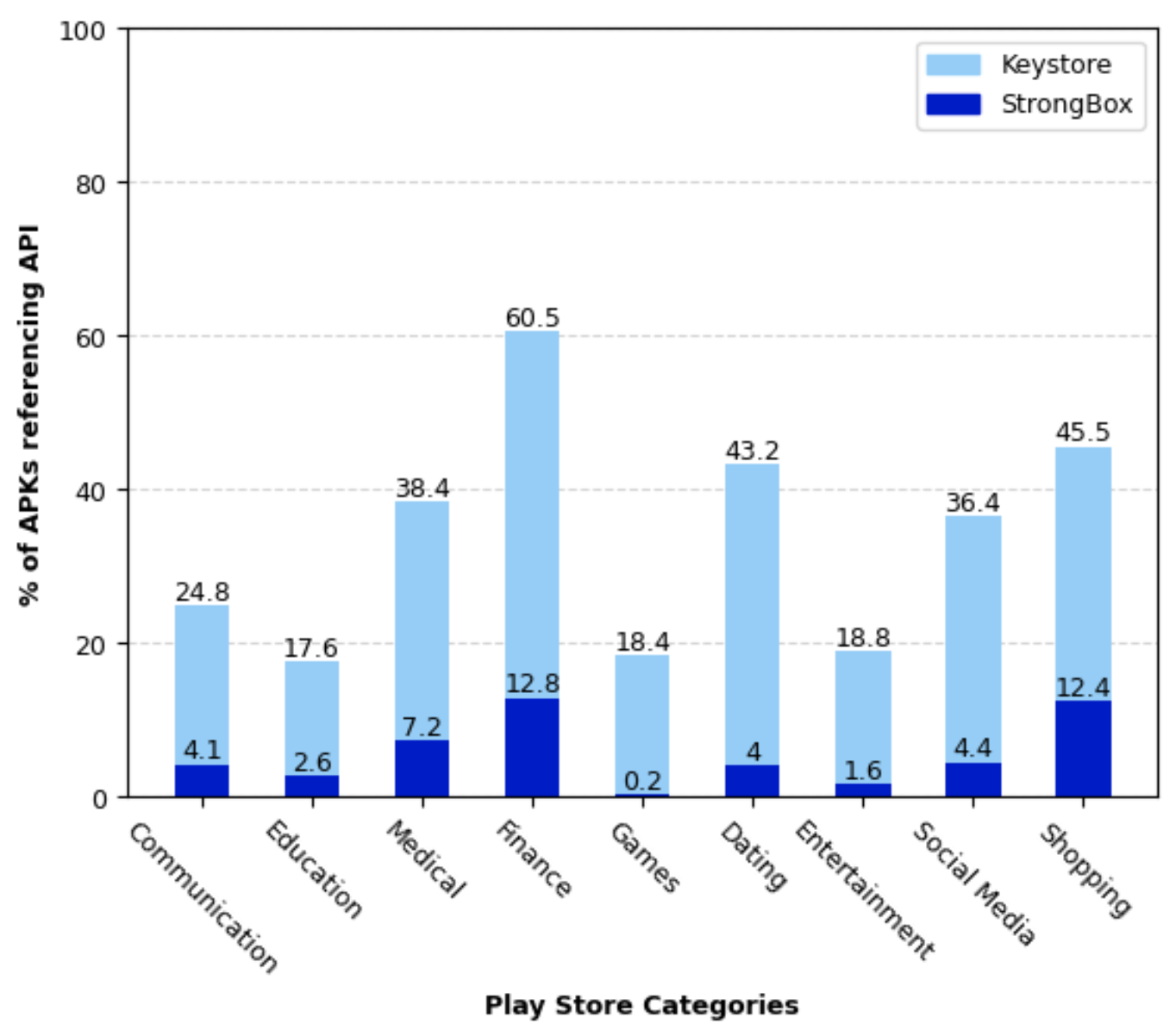}
    \caption{Percentages of Android apps using TEE and SE APIs, respectively, across major categories within the Google Play Store. StrongBox usage is shown here as a subset of Android Keystore API usage (i.e., any app that uses StrongBox necessarily uses the Android Keystore API).}
    \label{fig:category_usage}
\end{figure}

Figure~\ref{fig:category_usage} shows the comparative usage of the Android Keystore (TEE) and StrongBox (SE) APIs across a range of categories in the Google Play Store. We show a representative sample of categories here due to space limitations, but data for all categories can be found in the accompanying code repository (see~\S\ref{sec:dataset}). Unsurprisingly, we find that financial apps demonstrate the highest rates of trusted hardware usage, with over 60\% of apps referencing the broader Android Keystore API and 12.8\% referencing the StrongBox API. Gaming apps have an exceptionally low rate of StrongBox usage, which we hypothesize is due to the fact that the vast majority of StrongBox references come from third-party APIs, but gaming app development teams are less likely to use high-level app development toolkits given the more advanced functionality required to create the app.

\subsection{Manual Analysis}
\label{appendix:manual_analysis}
We select a subset of applications flagged as not using Android’s trusted hardware API for further examination to verify our static analysis results and to better understand which key storage schemes are used instead. We scraped the top 200 most-downloaded apps in the Play Store as of April 1, 2024, and then selected the ten most highly-ranked apps which self-reported collecting sensitive data but had been flagged in our initial keyword search as not referencing the Android Keystore API. We decompiled each app using Apktool and manually searched for relevant keywords relating to widely used software-backed keystores provided as part of Android, such as Android’s SharedPreferences API~\cite{android_sharedpreferences}.


We verified that each of these ten apps were indeed not using Android’s trusted hardware API anywhere and found that they instead generally made use of some combination of Android’s SharedPreferences, Android's default software-backed keystore (i.e., \texttt{AndroidOpenSSL}), or a local SQLite database such as SQLCipher. SharedPreferences is a bit more concerning than other software-backed keystores as it offers very different security properties: some Android keystores (namely \texttt{SharedPreferences} and \texttt{KeyChain} are intended as systemwide credential storage, where keys are accessible to \textit{any} app on the device. While it is challenging to make any definitive statements on individual app use cases due to the high-level nature of our analysis and obfuscation of internal variable names, developers should always exercise caution when using a systemwide keystore.

It is also possible that some apps hardcode encryption keys after obfuscating the keys using Dexguard or a similar tool, but this was not possible to detect given our manual review is relatively cursory and intended primarily to verify our static analysis results and identify other APIs used.

\subsection{Developer Survey Questions}
\label{sec:appendix_survey}

\begin{enumerate}
    \item{Which of the following best describes your role?}
        \subitem{a. Programmer/Developer}
        \subitem{b. Software Tester/Quality Assurance}
        \subitem{c. Project Manager}
        \subitem{d. Software Design/Architecture}
        \subitem{e. Administration (Non-Technical)}
        \subitem{f. Other}

\item{Approximately how many people (including project managers, developers, testers, etc.) are involved in developing the app?}
    \subitem{a. 1}
    \subitem{b. 2 - 5}
    \subitem{c. 6 - 20}
    \subitem{d. 21 - 50}
    \subitem{e. 50+}

\item{How many years of experience do you have working with Android app development?}
    \subitem{a. None}
    \subitem{b. Less than 1 year}
    \subitem{c. 1 - 5 years}
    \subitem{d. 5 - 10 years}
    \subitem{e. 10 years or more}

\item{On a five-point scale, how much do you agree with the following statement: Our development team prioritizes security as part of the development process.}
    \subitem{a. Strongly agree}
    \subitem{b. Agree}
    \subitem{c. Neither agree nor disagree}
    \subitem{d. Disagree}
    \subitem{e. Strongly disagree}

\item{On a five-point scale, how much do you agree with the following statement: Our app collects and processes potentially sensitive user data (e.g., name, other demographic information, health data, financial data, etc.).}
    \subitem{a. Strongly agree}
    \subitem{b. Agree}
    \subitem{c. Neither agree nor disagree}
    \subitem{d. Disagree}
    \subitem{e. Strongly disagree}

\item{On a five-point scale, how much do you agree with the following statement: I am familiar with the concept of trusted hardware (e.g., Intel SGX and Arm TrustZone).}
    \subitem{a. Strongly agree}
    \subitem{b. Agree}
    \subitem{c. Neither agree nor disagree}
    \subitem{d. Disagree}
    \subitem{e. Strongly disagree}

\item{On a five-point scale, how much do you agree with the following statement: I am familiar with the \href{https://developer.android.com/privacy-and-security/keystore}{Android Keystore} trusted hardware API, commonly used in Android development for credential storage (e.g., storing cryptographic keys).}
    \subitem{a. Strongly agree}
    \subitem{b. Agree}
    \subitem{c. Neither agree nor disagree}
    \subitem{d. Disagree}
    \subitem{e. Strongly disagree}

\item{[\textit{If app did not reference Android Keystore API at all.}] Based on our static analysis as of November 2023, your app was recorded as not using the Android Keystore API. Which of the following reasons best describe the main considerations behind this decision? Please select all that apply.}
    \begin{itemize}
        \item Security benefits were unclear
        \item Security benefits were not needed given type of data (if any) collected by app
        \item Performance concerns
        \item Lack of features: Desired algorithm and/or key size was unavailable with Android Keystore
        \item We wanted to maximize our app’s ability to run on many different devices (potentially running older versions of Android)
        \item App was developed prior to Android’s Keystore API release date in 2013
        \item Found Keystore API difficult to use
        \item Unaware this API existed
        \item Don’t know/don’t remember
        \item We believe your static analysis result to be incorrect: [open text]
        \item Other: [open text]
    \end{itemize}

\item{[\textit{If app did not reference Android Keystore API at all.}] To the best of your knowledge, what libraries, if any, does your app use within Android for credential storage (either user login credentials or developer credentials such as cryptographic keys)? [Open text]}

\item{[\textit{If app was recorded as disabling StrongBox.}] A secure element (called the \href{https://developer.android.com/privacy-and-security/keystore#HardwareSecurityModule}{StrongBox Keymaster} in Android) is a more advanced form of trusted hardware. Based on our static analysis of your app from November 2023, we determined your app used the Android Keystore API but disabled usage of the secure element in at least one instance. (Specifically, the setIsStrongBoxBacked API described in the link above was set to false). Which of the following reasons best describe the main considerations behind this decision? Please select all that apply.}
    \begin{itemize}
        \item Security benefits were unclear
        \item Security benefits were not needed given type of data (if any) collected by app
        \item Performance concerns
        \item Lack of features: desired cryptographic algorithm and/or key size was unavailable with StrongBox Keymaster
        \item We wanted to maximize our app’s ability to run on many different devices (potentially running older versions of Android)
        \item Don’t know/don’t remember
        \item We believe your static analysis result to be incorrect: [open text]
        \item Other: [open text]
    \end{itemize}

\end{enumerate}

\label{sec:api_totals}
\begin{table*}[h!]
\begin{tabular}{@{}lr@{}}
\toprule
\textbf{Keystore API Method}                             & \textbf{Count} \\ \midrule
void <init>(java.lang.String,int) & 278,567 \\ \hline
android.security.keystore.KeyGenParameterSpec\$Builder setEncryptionPaddings(java.lang.String[]) & 235,719 \\ \hline
android.security.keystore.KeyGenParameterSpec\$Builder setBlockModes(java.lang.String[]) & 224,169 \\ \hline
android.security.keystore.KeyGenParameterSpec\$Builder setKeySize(int) & 166,379 \\ \hline
android.security.keystore.KeyGenParameterSpec\$Builder setUserAuthenticationRequired(boolean) & 48,150 \\ \hline
android.security.keystore.KeyGenParameterSpec\$Builder setDigests(java.lang.String[]) & 48,095 \\ \hline
android.security.keystore.KeyGenParameterSpec\$Builder setCertificateNotAfter(java.util.Date) & 44,087 \\ \hline
android.security.keystore.KeyGenParameterSpec\$Builder setCertificateNotBefore(java.util.Date) & 44,062 \\ \hline
android.security.keystore.KeyGenParameterSpec\$Builder setRandomizedEncryptionRequired(boolean) & 30,245 \\ \hline
android.security.keystore.KeyGenParameterSpec\$Builder setIsStrongBoxBacked(boolean) & 24,656 \\ \hline
android.security.keystore.KeyGenParameterSpec\$Builder setUserAuthenticationValidityDurationSeconds(int) & 23,946 \\ \hline
android.security.keystore.KeyGenParameterSpec\$Builder setKeyValidityForOriginationEnd(java.util.Date) & 15,334 \\ \hline
android.security.keystore.KeyGenParameterSpec\$Builder setSignaturePaddings(java.lang.String[]) & 9,313 \\ \hline
android.security.keystore.KeyGenParameterSpec\$Builder setUserAuthenticationParameters(int,int) & 8,974 \\ \hline
android.security.keystore.KeyGenParameterSpec\$Builder setInvalidatedByBiometricEnrollment(boolean) & 6,629 \\ \hline
android.security.keystore.KeyGenParameterSpec\$Builder setAlgorithmParameterSpec\\(java.security.spec.AlgorithmParameterSpec) & 5,531 \\ \hline
android.security.keystore.KeyGenParameterSpec\$Builder setAttestationChallenge(byte[]) & 2,724 \\ \hline
android.security.keystore.KeyGenParameterSpec\$Builder setKeyValidityEnd(java.util.Date) & 1,295 \\ \hline
android.security.keystore.KeyGenParameterSpec\$Builder setKeyValidityStart(java.util.Date) & 1,088 \\ \hline
android.security.keystore.KeyGenParameterSpec\$Builder setUnlockedDeviceRequired(boolean) & 383 \\ \hline
android.security.keystore.KeyGenParameterSpec\$Builder setKeyValidityForConsumptionEnd(java.util.Date) & 230 \\ \hline
android.security.keystore.KeyGenParameterSpec\$Builder setUserAuthenticationValidWhileOnBody(boolean) & 93 \\ \hline
android.security.keystore.KeyGenParameterSpec\$Builder setUserConfirmationRequired(boolean) & 47 \\ \hline
android.security.keystore.KeyGenParameterSpec\$Builder setUserPresenceRequired(boolean) & 38 \\
\bottomrule
\end{tabular}

\caption{\textbf{Usage count of Android Keystore API methods across all apps in the Play Store.}}
\label{tab:keystore_api_refs}
\end{table*}


\begin{table}[t]
\begin{tabular}{@{}lR{1.7cm}}
\toprule
\textbf{Package Name}                             & \textbf{Call Count} \\ \midrule
com.google.android.gms.internal\\ \quad .firebase-auth-api & 30,055 \\ \hline
androidx.security.crypto & 26,345 \\ \hline
com.appsflyer & 23,566 \\ \hline
androidx.biometric & 15,960 \\ \hline
com.microsoft.appcenter.utils.crypto & 12,282 \\ \hline
com.google.crypto.tink.integration.android & 11,656 \\ \hline
com.flurry.sdk & 7,806 \\ \hline
com.amazonaws.internal.keyvaluestore & 4,138 \\ \hline
com.oblador.keychain.cipherStorage & 4,073 \\ \hline
com.huawei.secure.android.common \\ \quad .encrypt .keystore.aes & 2,794 \\

\bottomrule
\end{tabular}

\caption{Top 10 third-party libraries referencing the Android Keystore key initialization API \texttt{<init>(java.lang.String, int)}. We chose to classify \texttt{androidx.security.crypto}~\cite{androidx_crypto} as third party after finding that the majority of references were to the \texttt{EncryptedFile} and \texttt{EncryptedSharedPreferences} classes which abstract the details of key generation and storage.}
\label{tab:thirdparty_package_calls}
\end{table}


\begin{table}[t]
\begin{tabular}{@{}lR{1.7cm}}
\toprule
\textbf{Package Name}                             & \textbf{Call Count} \\ \midrule
androidx.security.crypto & 11,424 \\ \hline
com.oblador.keychain.cipherStorage & 2,161 \\ \hline
com.microsoft.identity.common.internal\\ \quad .platform & 1,019 \\ \hline
com.salesforce.marketingcloud.sfmcsdk \\ \quad .components.encryption & 758 \\ \hline
com.iproov.sdk.crypto & 179 \\ \hline
androidx.tracing & 136 \\ \hline
com.ionicframework.IdentityVault & 129 \\ \hline
com.oblador.keychain.g & 108 \\ \hline
com.epicshaggy.biometric & 76 \\ \hline
com.it\_nomads.fluttersecurestorage.ciphers & 66 \\

\bottomrule
\end{tabular}

\caption{Top 10 third-party libraries referencing Android's secure element StrongBox Keymaster API.. We chose to classify \texttt{androidx.security.crypto}~\cite{androidx_crypto} as a third-party library after finding that the majority of references were to the \texttt{EncryptedFile} and \texttt{EncryptedSharedPreferences} classes which abstract the details of key generation and storage from the developer.}
\label{tab:strongbox_thirdparty_package_calls}
\end{table}


\clearpage

\begin{table}[h]
\center
\begin{tabular}{@{}lR{3.0cm}}
\toprule
\textbf{Cipher}               & \textbf{Usage Count} \\ \midrule
AES                  & 147,529       \\
RSA                  & 80,096       \\
HMAC-SHA256          & 2,321         \\
EC                   & 2,233         \\
HMAC-SHA512          & 100          \\ \bottomrule
\end{tabular}

\vspace{2em}
\caption{List of ciphers requested for the Android Keystore provider through the \texttt{javax.crypto.KeyGenerator, java.security.KeyPairGenerator, javax.crypto.Cipher} and \texttt{java.security.KeyStore} APIs along with respective usage counts. We include results from both the ``AndroidKeyStore'' and ``AndroidKeyStoreBCWorkaround'' providers due to an Android bug dating back to 2015~\cite{android_bcworkaround}.}
\label{tab:keystore_ciphers}
\end{table}

\begin{table}[H]
\center
\begin{tabular}{@{}ccC{2.6cm}}
\toprule
\multicolumn{1}{c}{\multirow{2}{*}{\textbf{\begin{tabular}[c]{@{}c@{}}Message\\ Size (MiB)\end{tabular}}}} & \multicolumn{2}{c}{\textbf{Avg. Runtime (s)}} \\ \cmidrule(l){2-3} 
\multicolumn{1}{c}{}                                             & TEE                   & SE                    \\ \midrule
0.01                                                             & 0.03 $\pm$ 0.01                  & 0.21 $\pm$ 0.01                   \\
0.1                                                              & 0.08 $\pm$ 0.01                  & 1.59 $\pm$ 0.02                  \\
0.2                                                              & 0.12 $\pm$ 0.02                  & 3.11 $\pm$ 0.02                  \\
1                                                                & 0.42 $\pm$ 0.06                  & 15.43 $\pm$ 0.10                 \\
2                                                                & 1.13 $\pm$ 0.09                  & 30.88 $\pm$ 0.16                 \\
4                                                                & 2.56 $\pm$ 0.19                  & 62.23 $\pm$ 0.33                 \\
6                                                                & 4.25 $\pm$ 0.74                  & 94.68 $\pm$ 0.37                 \\
8                                                                & 7.67 $\pm$ 1.02                  & 159.61 $\pm$ 0.72                \\
10                                                               & 5.83 $\pm$ 0.63                  & 127.01 $\pm$ 0.69                \\
12                                                               & 9.24 $\pm$ 0.83                  & 192.37 $\pm$ 0.80                \\
14                                                               & 10.87 $\pm$ 1.14                 & 223.44 $\pm$ 1.02                \\
16                                                               & 13.10 $\pm$ 1.44                 & 257.69 $\pm$ 1.09                \\ \cmidrule(r){1-3}
\end{tabular}

\vspace{2em}
\caption{Execution times of AES-GCM-256 encryption as a function of message length. These measurements were taken from the TEE and SE in Google's Pixel 8 device.}
\label{tab:encryption_length_runtime}
\end{table}

\begin{table}[H]
\center
\begin{tabular}{@{}ccC{2.6cm}}
\toprule
\multicolumn{1}{c}{\multirow{2}{*}{\textbf{\begin{tabular}[c]{@{}c@{}}Message\\ Size (MiB)\end{tabular}}}} & \multicolumn{2}{c}{\textbf{Avg. Runtime (s)}} \\ \cmidrule(l){2-3} 
\multicolumn{1}{c}{}                                             & TEE                   & SE                    \\ \midrule
0.01                                                             & 0.02 $\pm$ 0.01                  & 0.15 $\pm$ 0.02                   \\
0.1                                                              & 0.06 $\pm$ 0.01                  & 0.99 $\pm$ 0.07                  \\
0.2                                                              & 0.14 $\pm$ 0.02                  & 1.89 $\pm$ 0.02                  \\
1                                                                & 0.48 $\pm$ 0.03                  & 9.05 $\pm$ 0.05                 \\
2                                                                & 0.89 $\pm$ 0.04                  & 17.99 $\pm$ 0.06                 \\
4                                                                & 1.76 $\pm$ 0.05                  & 35.91 $\pm$ 0.09                        \\ \cmidrule(r){1-3}
\end{tabular}

\vspace{2em}
\caption{Execution times of generating an Elliptic Curve Digital Signature Algorithm (ECDSA) signature with SHA-256. These measurements were taken from the TEE and SE in Google's Pixel 8 device.}
\label{tab:signing_length_runtime}
\end{table}


\end{document}